**Electrically and mechanically tunable electron spins in silicon carbide color centers**

Abram L. Falk[1,2], Paul V. Klimov[1,2], Bob B. Buckley[2], Viktor Ivády[3,4], Igor A. Abrikosov[4], Greg Calusine[2], William F. Koehl[1,2], Ádám Gali[3,5], and David D. Awschalom[1,2]

1. Institute for Molecular Engineering, University of Chicago, Chicago, IL 60637, USA
2. Center for Spintronics and Quantum Computation, University of California, Santa Barbara, Santa Barbara, CA 93106, USA
3. Wigner Research Centre for Physics, Institute for Solid State Physics and Optics, Hungarian Academy of Sciences, PO Box 49, H-1525, Budapest, Hungary
4. Dept. of Physics, Chemistry and Biology, Linköping University, SE-581 83 Linköping, Sweden
5. Dept. of Atomic Physics, Budapest University of Technology and Economics, Budafoki út 8., H-1111, Budapest, Hungary

**The electron spins of semiconductor defects can have complex interactions with their host, particularly in polar materials like SiC where electrical and mechanical variables are intertwined. By combining pulsed spin resonance with ab-initio simulations, we show that spin-spin interactions within SiC neutral divacancies give rise to spin states with an enhanced Stark effect, sub-$10^{-6}$ strain sensitivity, and highly spin-dependent photoluminescence with intensity contrasts of 15-36%. These results establish SiC color centers as compelling systems for sensing nanoscale fields.**

Leveraging advanced SiC processing capabilities alongside solid-state spin control is a promising pathway to quantum-information and sensing technologies [1-6]. Much like nitrogen-vacancy (NV) centers in diamond [7], neutral divacancies in SiC [1, 8-10] have a spin-triplet electronic ground state with optical addressability and long spin coherence times [4, 5] that persist up to room temperature. The NV center in diamond is currently being developed for applications ranging from quantum communication [11] to nanoscale nuclear magnetic resonance [12, 13]. The extensive uses of SiC in industry, including wafer-scale growth [14], high-power devices, and substrates for epitaxially grown GaN [15] and graphene [16], could propel these technologies and many others [17-19] forward. Underlying the success of such advancements will be an improved understanding of how spins in semiconductor defects interact with their host crystal.

In this letter, we demonstrate that the spin transitions of neutral divacancies in 4H-SiC are highly sensitive to electrical and mechanical perturbations of their host. They exhibit an electric-field response that is 2-7 times stronger than that [20, 21] for NV spins in diamond and highly spin-dependent photoluminescence (PL) for high-fidelity spin readout. Moreover, AC strain-sensing protocols demonstrated here lead to an optimized sensitivity projected to be in the $10^{-7}\ /\sqrt{\mathrm{Hz}\cdot N}$ range, where *N* is the number of interrogated spins. Although electric- and strain-field effects on SiC spins are interrelated due to SiC's piezoelectricity, our *ab-initio* simulations disentangle these two effects and attribute the strong electric field response of these spins to the high electron polarizability in SiC. Because our techniques are based on point defects and measure intrinsic quantities, their applicability extends down to the nanometer scale.

Neutral divacancies consist of neighboring C and Si vacancies and exist in 4 inequivalent forms in 4H-SiC, which we label PL1-PL4. Their ground-state triplet spin states are described by the Hamiltonian [20, 22]:

$$H = hD\sigma_z^2 + g\mu_B\boldsymbol{\sigma}\cdot\mathbf{B} - E_x\left(\sigma_x^2 - \sigma_y^2\right) + E_y(\sigma_x\sigma_y + \sigma_y\sigma_x), \qquad (1)$$

where the defect axis is aligned along **z**, *h* is Planck's constant, *g* = 2.0 is the electron g-factor, $\mu_B$ is the Bohr Magneton, **σ** is the vector of spin-1 Pauli matrices, **B** is the magnetic field, and *D,* $E_x$, and $E_y$ are the zero-magnetic-field splitting parameters. These terms can be expanded as $D = D^0 + d_{||}F_z + e_{||}\varepsilon_z$ and $E_{x,y} = E_{x,y}^0 + d_\perp F_{x,y} + e_\perp \varepsilon_{x,y}$, where the $D^0$ and $E_{x,y}^0$ terms are the crystal-field splittings in the absence of applied strain and electric fields, $d_{||}$ and $d_\perp$ are the Stark-coupling parameters of the ground-state spin to an electric field (**F**) that is respectively parallel and perpendicular to the defect axis, $e_{||}$ and $e_\perp$ are the strain-coupling parameters, and **ε** is the effective strain field defined in Ref. [22]. PL1 and PL2 are oriented along the SiC c-axis and have $C_{3v}$ symmetry ($E_{x,y}^0$=0). PL3 and PL4 are oriented along basal planes, at 109.5° from the c-axis. This orientation reduces their symmetry to $C_{1h}$, resulting in nonzero $E_{x,y}^0$ and thus broken degeneracy between all three spin sublevels at zero magnetic field.

Our experiments use high-purity semi-insulating 4H-SiC wafers, purchased from Cree Inc., in which neutral divacancies are incorporated during crystal growth. We thinned 500 μm-thick chips of SiC down to 50 μm-thick membranes and used epoxy to mount them on top of piezo actuators, which are in turn mounted to copper cold fingers for cryogenic operation. The neutral divacancies' zero-phonon line (ZPL) optical transitions can be seen as peaks in their PL spectra [10] when we illuminate a SiC membrane with 1.27 eV light at a temperature (*T*) of 20 K. Each inequivalent divacancy has a distinct ZPL energy ranging from 1.0-1.2 eV. In addition, two other observed defects (labelled PL5 and PL6) have similar optical and spin transition energies to the neutral divacancies [4], but the defects with which they are associated have not been identified. Microwave radiation for electron spin resonance is supplied by waveguide antennae on chip or below the sample (see Refs. [4, 5] for experimental details). The piezo actuation applies tensile or compressive strain to the SiC membrane normal to the c axis (Fig. 1a), which we estimate to be $5 \times 10^{-7}$ strain/$V_{\mathrm{piezo}}$, where $V_{\mathrm{piezo}}$ is the voltage applied to the piezo (see supplementary information (SI)). Control measurements assure that electric fields within the measurement volume due to $V_{\mathrm{piezo}}$ are too weak to interfere with any measurements.

As strain is applied, the energy of each defect species' ZPL splits and shifts as much as 2.3 meV, or 550 GHz (Fig. 1b). The ZPLs corresponding to the c-axis-oriented defects (PL1, PL2, and PL6) bifurcate, with the two resulting branches having orthogonally polarized PL (Fig. 1c). This splitting reflects the reduction of the $C_{3v}$ symmetry of the c-axis-oriented defects, whose doubly-degenerate excited state orbitals at zero strain are predicted [1] to closely match the structure of NV centers in diamond [23, 24]. In contrast, the basal-oriented defects (PL3, PL4, and PL5) have highly split ZPLs even at zero strain due to the crystal field. Each ZPL branch of the basal-plane-oriented defects trifurcates as strain is applied, with the polarization from each branch offset by 120°. This splitting indicates that, as expected, strain breaks the symmetry between orientations of defects in the basal plane that are equivalent at zero strain.

Applying transverse strain to the SiC membrane also shifts the defect's electronic spin transition energies. We measure these shifts with optically detected magnetic resonance (ODMR). Here, the ground-state spin is read out by exploiting the defects' spin-dependent ($m_s = 0$ vs. $m_s = \pm 1$) PL intensity ($I$) and monitoring the changes in $I$ ($\Delta I$) as the spins are rotated between spin eigenstates via electron spin resonance [4, 5].

The piezo actuator is found primarily to shift the two $\Delta m_s = \pm 1$ transitions of the c-axis-oriented divacancies together (Fig. 2a), as opposed to splitting them, indicating that the applied uniaxial stress primarily affects the $D$ term of the Hamiltonian (Eq. 1) in this measurement. Although we observe DC spin resonance shifts of up to 0.8 MHz, the resolution of DC strain detection [22] is constrained by relatively broad $1/T_2^*$-limited spin-resonance linewidths, where $T_2^*$ is the inhomogeneous ensemble dephasing time (~1.5 μs for these defects).

Our AC- strain-sensing technique enables more sensitive strain measurements that are limited by the much longer ensemble homogeneous coherence time ($T_2$), which is up to 360 μs for divacancies in SiC at 20 K [5]. We use Hahn-echo-based sensing [5, 20] to measure the change in the spins' precession rate between $m_s$=0 and $m_s$=±1 spin states due to a synchronized $V_{piezo}$ waveform (Fig. 2b). The difference in strain applied before and after the central π pulse of the sequence causes a change in the spin precession rate across the two halves of the sequence (see SI for details). In turn, this time-integrated difference phase-shifts the superposition of the spin echo. As $V_{piezo}$ is increased, $\Delta I/\Delta I_{Hahn}$ oscillates according to this strain-induced phase shift (Fig 2c), where $\Delta I_{Hahn}$ is the signal strength from a Hahn-echo sequence without $V_{piezo}$ applied.

Despite the large uncertainty in our strain calibration (40%, see SI), we infer that the spin transitions of PL1-PL4 exhibit strain shifts roughly ranging from 2-4 GHz/strain, and that our ensemble measurements demonstrate a sub-$10^{-6}$ strain sensitivity after averaging for two minutes per point. In an ideal measurement, with high optical collection and small background PL, AC strain sensing with neutral divacancies has a projected sensitivity in the $10^{-7}/\sqrt{\mathrm{Hz} \cdot N}$ range (see SI).

To measure the spin states' response to electric fields, we use the same pulse sequence as for strain measurements (Fig. 2b), except that we apply a voltage ($V_{electrode}$) to electrodes across the SiC membrane (parallel to the crystal c-axis) instead of applying piezo-actuated strain (Fig. 3a) [20]. The basal-oriented defect spins (PL3-PL5) primarily couple to the c-axis electric field via $d_\perp$, causing their two $\Delta m_s$=±1 spin transition energies to shift in opposite directions. The $\Delta I$ signals from the two spin-resonance transitions in our pulse sequence are therefore out of phase (Fig. 3b). The c-axis-oriented defects (PL1, PL2, and PL6) have slower $\Delta I$ oscillations, indicating that their $d_{||}$ parameter (Table 1) is significantly smaller than $d_\perp$ for the basal-oriented defects.

An important parameter for high-fidelity sensing is the ODMR contrast ($C_{defect}$) between spin states, defined as the fractional change in PL due to an optically polarized spin population being flipped by $\Delta m_s$=±1. We use spectrally resolved ODMR measurements to measure $C_{defect}$ from ZPLs, avoiding interference from background PL (Fig. 4a). Instead of applying ideal π pulses to flip spin states, we mix the spin populations with strong continuously applied microwave radiation and weak laser illumination. Calculating $C_{defect}$ from the measured fractional change in PL intensity ($\Delta I_{mixed}/I$) requires detailed knowledge of spin mixing dynamics and rates, but use 3/2 × $\Delta I_{mixed}/I$ as a conservative lower bound for $C_{defect}$ (see SI). The inferred $C_{defect}$ lower bounds are found to range from 0.15-0.21 for the neutral divacancies (PL1-PL4), and from 0.33-.36 for PL5 and PL6 at 20 K. These high $C_{defect}$ values are comparable to that for NV centers in diamond, whose optimized ODMR contrast is typically around 0.3 [25, 26]. The optical lifetimes of the SiC divacancies, another important quantity for spin readout, are found to range from 12-15 ns (Fig. 4b and SI), also comparable to that for NV centers in diamond [25]. These favorable ODMR characteristics make SiC defects compelling systems for precision sensing.

In order to understand the interplay between strain- and electric field-induced spin shifts in SiC, we carry out *ab-initio* density-functional calculations to calculate the zero-field splitting parameters in the ground-state Hamiltonian, along with electric- and strain-field perturbations to them. Our simulations calculate spin-spin interactions between electrons localized to the defects but neglect spin-orbit coupling. We therefore focus on the c-axis-oriented divacancies (PL1 and PL2), whose $C_{3v}$ symmetry implies low spin-orbit coupling. Our simulation methodology (see SI) provides excellent agreement with the ground state $D^0$ values for PL1 and PL2 in 4H-SiC, as well as for the NV center in diamond (within 1.5%). As a corroboration for the simulation methods and parameters, it also accurately calculates the change in zero-field splitting due to external pressure acting on NV center in diamond (12.6 MHz/GPa calculated vs. 14.6 MHz/GPa experimental [22].)

The experimental observation that compressive and tensile strain perpendicular to the c axis primarily shifts the *D* term of the Hamiltonian for the c-axis oriented defects (see Fig. 2a) is supported by simulations. These show that this strain causes only a small deviation from $C_{3v}$ symmetry for the defect's electronic orbitals, with their spin exhibiting a correspondingly small $E_{x,y}$ term (see SI), and *D* shifting by 7 GHz / strain, neglecting Poisson effects. These magnitudes are comparable to the experimentally determined (2-4 GHz/strain) values, but both are only order-of-magnitude estimates. For c-axis-oriented-strain, a *D* shift of 5 GHz / strain is calculated.

The measured electric field-spin coupling coefficients are consistently found to be higher than those [20, 21] for the diamond NV center, up to 1.9x higher for $d_{\perp}$ and 7.6x higher for $d_{||}$ (Table 1). Our simulations overestimate $d_{||}$ for both neutral divacancies in SiC and the diamond NV center, but they corroborate the numerical values of relative enhancement of $d_{||}$ for neutral divacancies in SiC over that for diamond NV centers. The simulated and experimentally derived relative enhancements agree within 30%.

Two physical effects contribute to $d_{||}$. First, electric fields distort the positions of atoms in the SiC lattice neighboring the divacancy. Second, electric fields shift the electron distribution surrounding the defect. Both effects influence the spin-density matrix of the system, shifting the spin transition energies in the ground-state spin Hamiltonian (Eq. 1). Although these effects occur simultaneously and cannot be distinguished by experiments alone, simulations can separate the two effects.

The piezoelectric effect primarily shifts the coordinates of the C atoms closest to the Si-vacancy portion of the divacancy. By calculating $d_{||}$ with a distorted lattice but no extra electric-field-induced shifts to electron wavefunctions, we find $d_{||}$ to be an order of magnitude smaller than its value when electric fields are turned on (Table 1). Thus, direct shifts of electron wavefunctions by external electric fields are primarily responsible for the Stark-shift parameters, not the piezoelectric effect.

The enhanced Stark effect of divacancy spins in SiC over NV-center spins in diamond can be understood by differences in electron polarizability. The polar crystal bonds in SiC result in high electron bond polarizability, which can be seen by the material's high (10.0) dielectric constant. Since defect wavefunctions are derived from bond orbitals, the dangling bonds that form the defect state also exhibit high polarizability (see Fig S10 in SI). In turn, this high polarizability causes a strong spin response to external electric fields.

Our results show that the spins of neutral divacancies in SiC can sensitively detect both strain and electric fields. They have high optical polarization [5], high intrinsic ODMR contrasts, and a stronger response to electric fields than those of NV centers in diamond. In the future, basal-oriented-defect spins in SiC could be ideal for temperature sensing [18, 19, 27-29], since their large transverse crystal-field splitting gives them first-order insensitivity to magnetic fields. Moreover, the combination of the spin-strain interactions measured here and the outstanding electromechanical properties of SiC [30] could make SiC an ideal material for coupling spins to mechanical resonators [31]. These coupled systems could lead to mechanically-induced spin squeezing [32], strong coupling between spins and phonons, and phonon lasing [33].

**Acknowledgments:** The authors thank Martin Stevens, Sae Woo Nam and Richard Mirin at NIST in Boulder, Colorado for providing assistance measuring the optical lifetimes. This work is supported by the AFOSR, DARPA, and NSF. VI, IAA and AG acknowledge the support from Knut and Alice Wallenberg Foundation ( "Isotopic Control for Ultimate Material Properties" project) and the National Supercomputer Center (SNIC 001-12-275). AG thanks the support from EU Commission (FP7 Grant No. 270197 - DIAMANT).

## Figures and tables

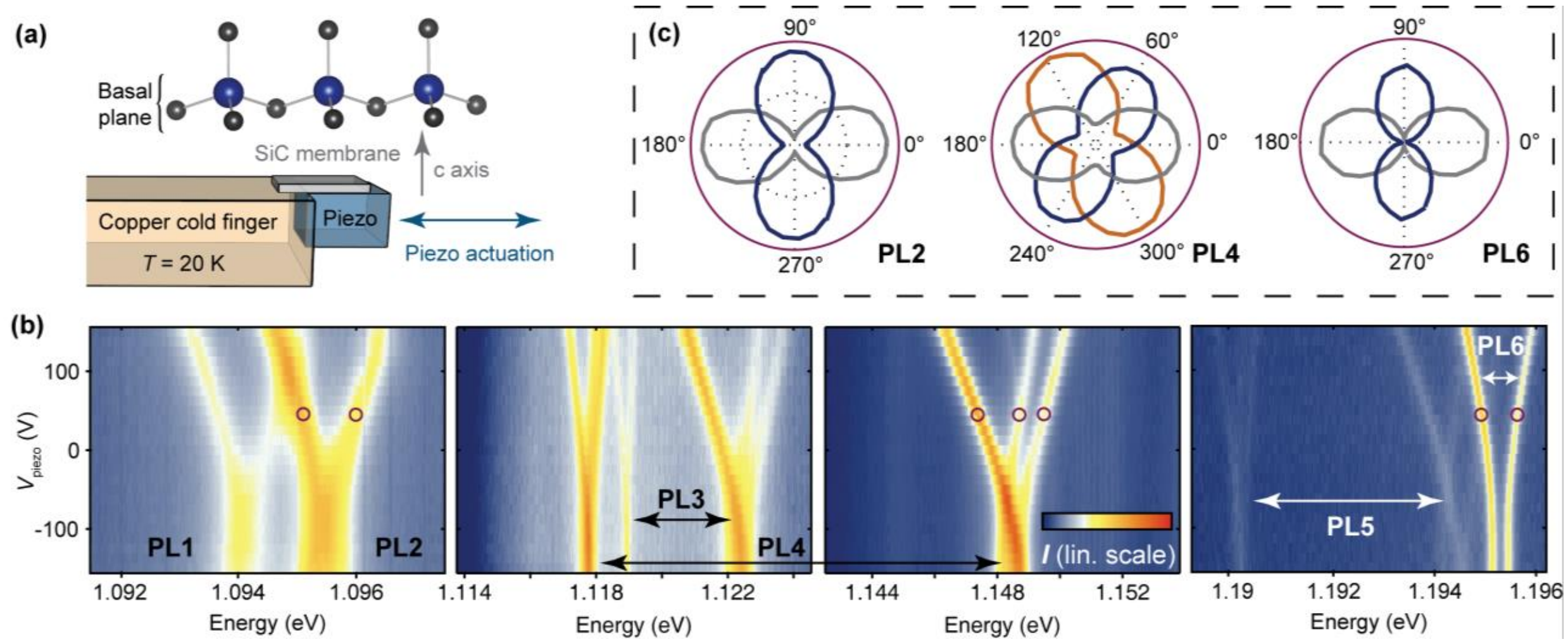

**FIG. 1. (a)** A SiC membrane is epoxied to the top of a piezo actuator, which applies strain to the SiC membrane as it stretches. **(b)** PL spectra as a function of $V_{\mathrm{piezo}}$ (strain) at $T$=20 K, showing that the optical transitions of SiC defects can be tuned with strain. The applied strain splits the ZPL optical transitions, with the c-axis-oriented defects (PL1, PL2, and PL6) bifurcating and the basal-oriented defects (PL3, PL4, and PL5) trifurcating. **(c)** Polarization dependence of the PL from the strain-split ZPL branches, measured at the points indicated by purple circles in (b). The analyzed polarization is in the plane perpendicular to the c-axis.

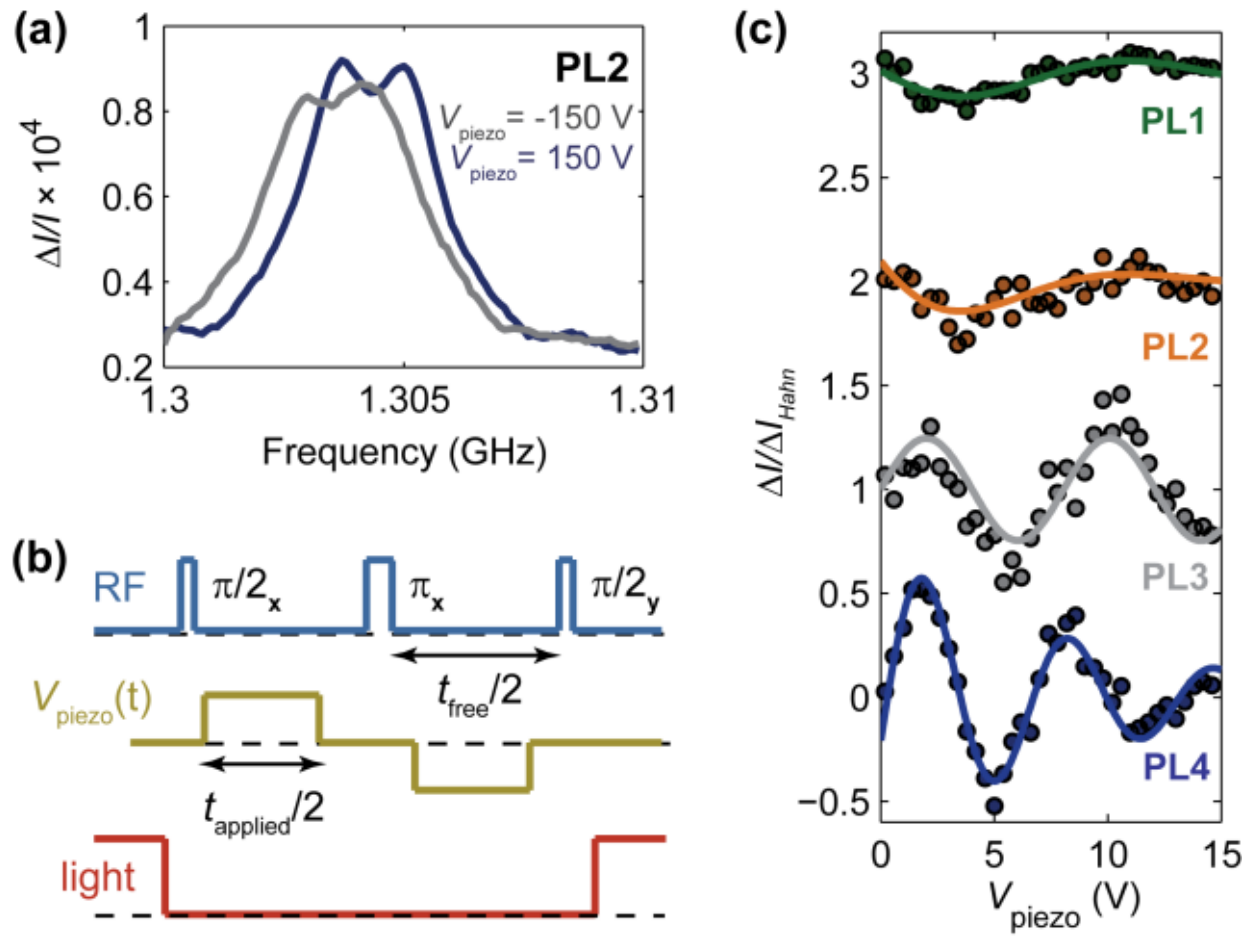

**FIG. 2.** (**a**) An ODMR measurement of the 0.8 MHz strain-induced shift in $D$ for PL2. The split peak within each curve is due to a nonzero stray $B$ field of 0.35 G. **(b)** The modified Hahn-echo pulse scheme that we use for AC strain sensing. (**c)** AC strain sensing data for PL1-PL4, with $t_{\mathrm{free}}$=100 μs, $t_{\mathrm{applied}}$= 80 μs, $T$=20 K, and $B$=0. The fits are to single exponentially decaying sinusoids and have frequencies of 0.07 $V^{-1}$, 0.07 $V^{-1}$, 0.12$V^{-1}$, and 0.16 $V^{-1}$ for PL1-PL4 respectively. The three curves are offset from the origin for clarity.

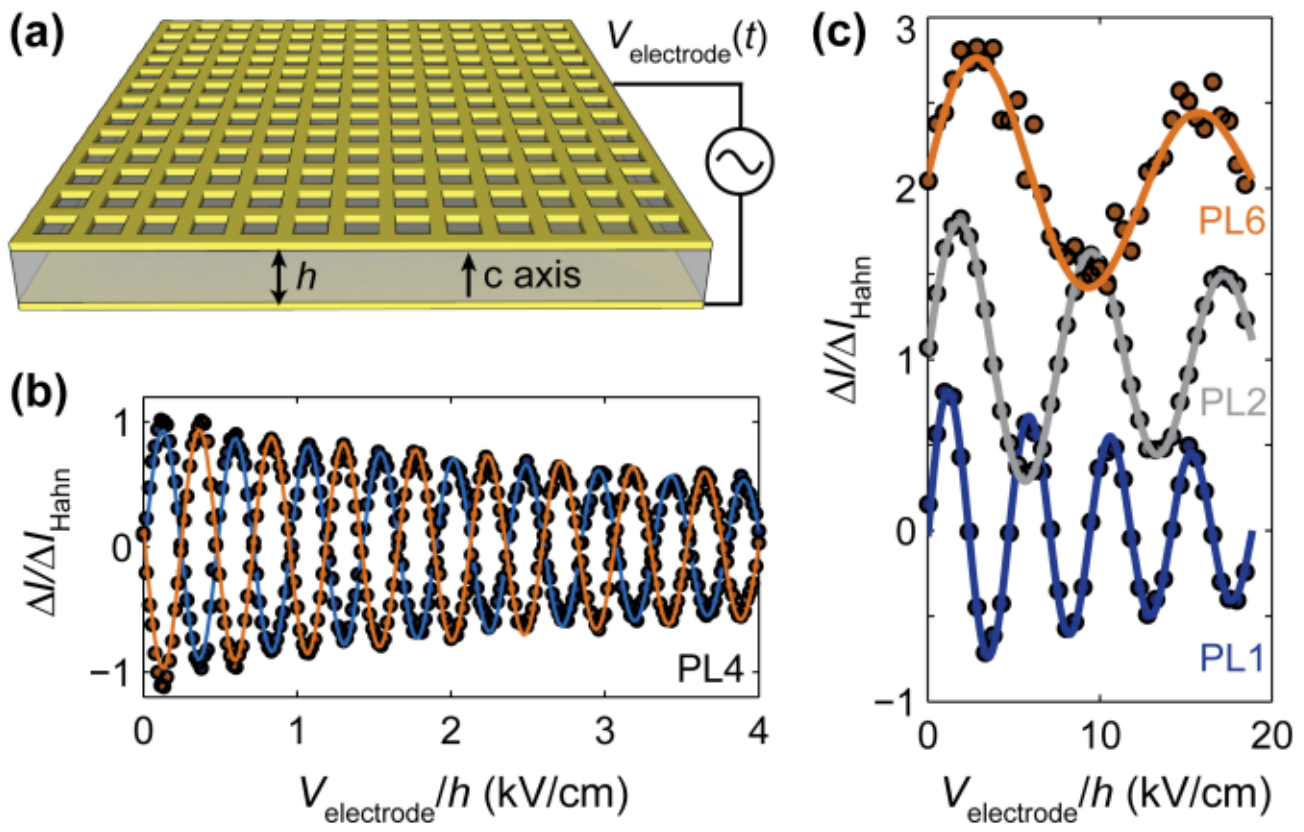

**FIG. 3. (a)** To sense electric fields with divacancy spins in SiC, $V_{electrode}$ is applied across an $h$ = 53 μm thick SiC membrane with electrodes. The top electrode is a patterned grating, allowing light to be transmitted. The pulse pattern is the same as that used in Fig. 2b-c, except that $V_{electrode}(t)$ is substituted for $V_{piezo}(t)$. (**b**) AC electric field-sensing measurements for basal-oriented PL4, using both the higher (blue) and lower (orange) frequency spin-resonance transitions. **(c)** Electric field sensing for PL1, PL2, and PL6, the c-axis oriented defects. All data are taken at $T$ = 20 K, $t_{applied}$ = 80 μs, $t_{free}$ = 100 μs, and $B$ = 0. The fits are to exponentially decaying sinusoids, with the three curves offset from the origin for clarity.

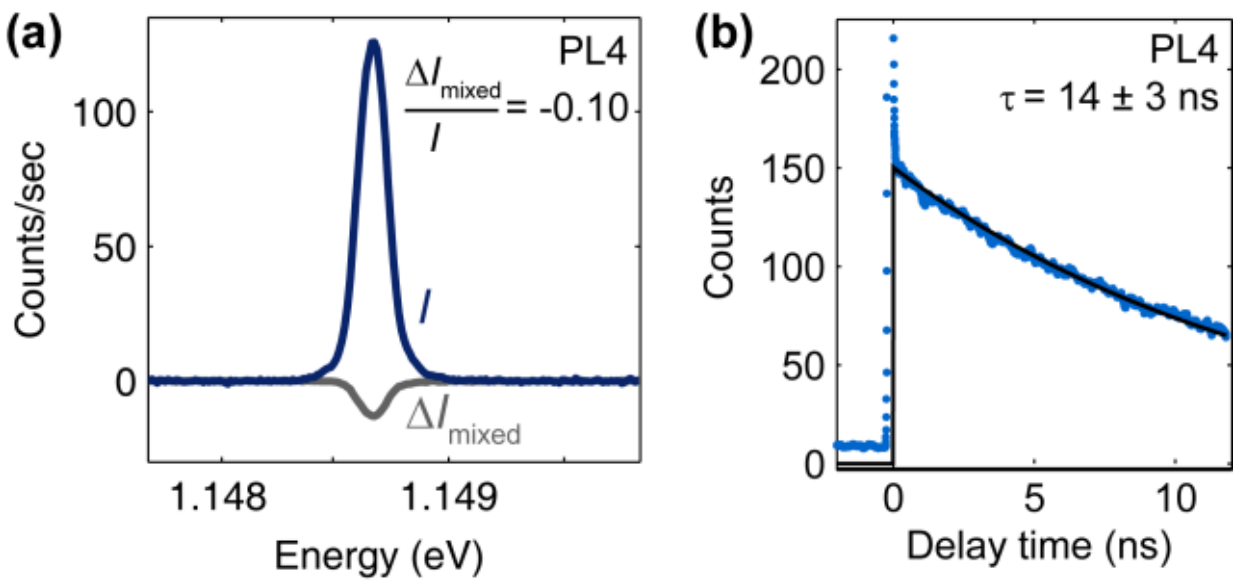

**FIG. 4. (a)** Plot of the ZPL intensity of PL4 without microwave radiation (blue curve) and the change in PL when strong microwave irradiation is applied to the sample (grey curve). $\Delta I_{mixed}/I$ is measured to be -0.10, +0.11, -0.14, -0.10, -0.24, and -0.22 for PL1-PL6 respectively. **(b)** Time-resolved PL from PL4 as a function of delay from an excitation laser pulse. The decay constant of the exponential fit (black) is the optical lifetime (τ). $T$ = 20 K, $B$ = 0. See the SI for the calculation of $C_{defect}$ from $\Delta I_{mixed}/I$ and for the other defect species' lifetimes and ODMR contrasts.

| **Defect/ Configuration** | **Experiment** | | **Theory** | |
| --- | --- | --- | --- | --- |
| | $d_{\perp}$ | $d_{\|\|}$ | $d_{\|\|}$ | $d_{\|\|}$ (*) |
| **NV center** | 17 | 0.35 | 0.76 | |
| **PL1** | | 2.65 | 5.2 | 0.38 |
| **PL2** | | 1.61 | 4.2 | 0.23 |
| **Ratio of PL1:NV** | | **7.6:1** | **6.8:1** | **0.5:1** |
| **Ratio of PL2:NV** | | **4.6:1** | **5.5:1** | **0.3:1** |
| **PL3** | 32.3 | < 3 | 0.41 | |
| **PL4** | 28.5 | 0.44 | 0.79 | |
| **PL5** | 32.5 | < 3 | | |
| **PL6** | | 0.96 | | |

(*) Calculated with only atom-distortion effect

**TABLE 1.** Experimentally measured and calculated Stark effect parameters for the PL1-PL6 ground-state-spin Hamiltonian in 4H-SiC, in units of $h$·Hz cm/V, and a comparison to those for the NV center in diamond, with Ref [21] used as an experimental reference for the diamond NV center. The experimental (calculated) values of $d_{||}$ are compared to the experimental (calculated) value for the diamond NV center (in bold). The experimental uncertainty for the SiC data is 5% (see SI).